\documentclass[sigconf,nonacm]{acmart}

\usepackage{amsmath,amsfonts}
\usepackage{algorithmic}
\usepackage{caption}
\usepackage{graphicx}
\usepackage{subcaption}
\usepackage{textcomp}
\usepackage{xcolor}
\usepackage{url}
\usepackage{adjustbox}
\usepackage{multirow}
\usepackage{enumitem}
\usepackage[linesnumbered,ruled]{algorithm2e}
\usepackage{bm}
\usepackage{setspace}
\usepackage{pifont}
\usepackage{xspace}

\usepackage{flushend}
\usepackage{balance}
\usepackage{fancyhdr}

\newcommand{\sysname}{ITME\xspace}

\setcopyright{none}
\renewcommand\footnotetextcopyrightpermission[1]{}

\makeatletter
\def\shorttitle{}
\def\shortauthors{}
\makeatother

\begin{document}

\title{ITME: Inference Tiered Memory Expansion with Disaggregated CXL-Hybrid Memories}

\author{Hakbeom Jang, Younghoon Min, Sunwoong Kim, Taeyoung Ahn, Hanyee Kim, \\ Youngpyo Joo, Hoshik Kim, and Jongryool Kim}
\affiliation{%
  \institution{\Large Memory Systems Research, SK hynix}
  \country{}
}

\begin{abstract}
The rapid shift toward agentic and long-context workloads in Large Language Models (LLMs) is pushing the industry beyond the capacity of individual servers toward disaggregated shared storage to handle TB-scale context states. This movement has led to the emergence of specialized shared context layers designed to externalize and share cumulative inference states across distributed clusters. While offloading to a data processing unit (DPU) within just-a-bunch-of-flash (JBOF) architectures accelerates NVMe-over-fabrics (NVMe-oF) target processing, the need for sophisticated software-level optimization and cost-efficiency burdens remain significant. Consequently, the ideal architecture for scaling this shared context infrastructure is still an active area of exploration.

In this paper, we propose \sysname{} (Inference Tiered Memory Expansion), which leverages a CXL-hybrid memory to present a massive, TB-scale byte-addressable remote memory expansion. This approach enables cost-efficient scaling and simplifies the software stack through direct byte-addressability, effectively addressing the challenges of shared context infrastructure. Our key insight is that the deterministic access patterns of voluminous model weights and prefix caches enable the system to proactively manage data movement across the memory-storage hierarchy. Leveraging this predictability, \sysname{} implements a pipelined, multi-tier DMA-based prefetching that orchestrates seamless data movement directly from the CXL-hybrid memory device to GPU memory. We validate \sysname{} by evaluating its performance potential with production-grade SK Hynix CMM and PCIe Gen5 NVMe SSDs, while further demonstrating its functional feasibility through an FPGA-based hardware prototype. Overall, \sysname{} enhances conventional CPU-offloading by providing additional remote memory expansion to accommodate large KV cache footprints beyond host memory limits, achieving up to a 35.7\% throughput improvement.
\end{abstract}

\maketitle

\fancypagestyle{standardpagestyle}{%
  \fancyhf{}%
  \fancyfoot[C]{\thepage}%
  \renewcommand{\headrulewidth}{0pt}%
  \renewcommand{\footrulewidth}{0pt}%
}

\fancypagestyle{firstpagestyle}{%
  \fancyhf{}%
  \fancyfoot[C]{\thepage}%
  \renewcommand{\headrulewidth}{0pt}%
  \renewcommand{\footrulewidth}{0pt}%
}

\fancypagestyle{plain}{%
  \fancyhf{}%
  \fancyfoot[C]{\thepage}%
  \renewcommand{\headrulewidth}{0pt}%
  \renewcommand{\footrulewidth}{0pt}%
}

\pagestyle{standardpagestyle}
\thispagestyle{firstpagestyle}

\section{Introduction}

As Large Language Models (LLMs) continue to grow in scale, the primary bottleneck in inference systems has shifted from raw compute performance to memory capacity~\cite{kwon2023vllm, cheng2025lmcache, qin2025mooncake, sheng2023flexgen, infinigen-osdi-2024}. Modern foundation models, such as OPT-175B, require hundreds of gigabytes of memory simply to store model weights (e.g., 325 GB in FP16), making it challenging to fit within the memory of a single high-end GPU~\cite{sheng2023flexgen}. At the same time, the rise of agentic AI workflows and long-context applications has dramatically increased the importance of the key-value (KV) cache, transforming it from a transient runtime buffer into a long-lived inference state that may need to persist across multiple turns and sessions~\cite{gao2024cachedattention}. These trends burden inference infrastructure by requiring the concurrent management of high-capacity model parameters and ever-growing runtime context data across memory hierarchies.

Expanding high-speed memory capacity, however, faces fundamental physical and economic constraints. GPU memory, such as HBM, can typically be increased only by adding more GPUs, which is both costly and often inefficient. Similarly, host memory capacity remains limited by CPU-dependent factors such as socket count and memory channel availability~\cite{wang-cxl-realworld-arxiv24}. As a result, modern inference systems rely on a multi-tier hierarchy spanning from capacity-constrained GPU memory to elastic remote storage~\cite{cheng2025lmcache, qin2025mooncake, aminabadi2022deepspeed}.

Recently, the conventional inference hierarchy has begun to evolve beyond local storage toward a more centralized and reusable design. In large-scale serving systems, keeping KV cache state in local SSDs leads to fragmentation across nodes and limits reuse when requests are rescheduled or migrated. To address this limitation, recent systems like NVIDIA CMX Context Memory Storage~\cite{nvidia_icms_2026} have introduced a disaggregated, intermediate shared storage tier for inference states. This disaggregated context storage layer enables multiple compute nodes in a cluster to access a shared KV cache pool through high-speed interconnects such as RDMA. This shift improves resource utilization and enables context reuse at cluster scale, making disaggregated inference increasingly practical for long-context and multi-turn workloads.

To deploy such a disaggregated storage tier at scale, data centers increasingly adopt high-density, energy-efficient architectures, such as DPU-based just a bunch of flash (JBOF) systems~\cite{sun2025scalio, guo2023leed, supermicro2024jbof, nvidia2021bluefield, chelsio2024t7dpu}. By design, these JBOF nodes prioritize massive capacity and cost-effectiveness by strictly limiting local CPU and memory resources. Modern DPUs leverage a specialized hardware optimization known as NVMe-over-fabrics (NVMe-oF) target offload~\cite{sun2025scalio}. This architecture delegates the handling of SSD operations entirely to the host channel adapter (HCA) hardware via PCIe peer-to-peer (P2P) communication. By bypassing the host CPU, this offloading technique drastically minimizes computational overhead, allowing the JBOF system to scale effectively and handle higher IOPS with lower latency. While DPU-based offloading resolves the local CPU bottleneck, simply scaling out these expensive devices is not cost-effective ~\cite{shinatc2020}; thus, various software techniques have been proposed to maximize their efficiency~\cite{sun2025scalio, guo2023leed, min2021gimbal, zhan2024tricklekv}. Despite these advancements, there is currently no clear consensus on the optimal way to handle the massive KV cache state required to realize a shared storage tier for large-scale LLM inference.

In this paper, we propose \sysname{} (Inference Tiered Memory Expansion),which leverages a CXL-hybrid memory to provide cost-efficient, TB-scale memory expansion for large-scale LLM workloads. The remote region functions as a massive, byte-addressable memory, providing GPU servers with a scalable extension of their physical memory via standard RDMA protocols. To maximize the efficiency of this tiered memory architecture, \sysname{} strategically targets LLM data types that exhibit high predictability and dominate the overall memory footprint. Table\ref{tab:data_analysis} summarizes the characteristics and target tiering for LLM data types. Activations transient and small with high performance criticality. Working KV caches, although larger than activations, are also latency critical, while having lower predictability. Consequently, activations and working KV cache are better placed in high-speed GPU or host memory tiers. In contrast, model weights and long-context KV caches, which represent the vast majority of TB-scale LLM workloads, are characterized by deterministic or moderate predictability and their large capacity. Leveraging this observation, \sysname{} offloads these voluminous, predictable data types to the remote expansion tier, effectively masking access latencies through proactive placement.
 
High predictability of these data types allows \sysname{} to effectively hide both storage and network overheads through a coordinated HW/SW prefetching strategy. Internal storage latencies are overcome by hardware-level prefetching from NVMe SSDs into the internal DRAM cache of the CXL-hybrid-memory, while network communication overheads are mitigated via a pipelined, multi-tier DMA-based prefetching mechanism that orchestrates data transfers from \sysname{} to the GPU via RDMA and host CPU memory. By overlapping this end-to-end data pipeline with GPU computation, \sysname{} effectively masks total latency, enabling seamless TB-scale memory expansion for LLM inference.

\begin{table}[t]
\centering
\caption{LLM Data Characteristics and Target Tiering}
\label{tab:data_analysis}
\resizebox{\columnwidth}{!}{%
\begin{tabular}{|l|c|c|c|c|}
\hline
\textbf{Data Type} & \textbf{Predict.} & \textbf{Perf.} & \textbf{Target Tier} & \textbf{Capacity} \\ \hline
Weights            & Deterministic      & Critical       & Our (\sysname{})            & Static (Large)     \\ \hline
Activations        & High               & Critical       & GPU Mem. / Host Mem.        & Transient (Small)  \\ \hline
Working KV         & Low                & Critical       & GPU Mem. / Host Mem.        & Incremental (Moderate) \\ \hline
Long-context KV    & Moderate           & Moderate       & Our (\sysname{})            & Cumulative (Large) \\ \hline
\end{tabular}%
}
\end{table}

The main contributions of this work are:
\begin{itemize}
\item \textbf{ITME: Inference Tiered Memory Expansion:} We propose \sysname{}, a tiered memory expansion architecture based on CXL-hybrid-memory for TB-scale LLM workloads. By organizing remote resources into a byte-addressable memory, \sysname{} enables GPU servers to access massive model weights and KV caches via standard RDMA, effectively extending memory capacity.

\item \textbf{CXL-Hybrid Memory Architecture:} We design a CXL-hybrid-memory architecture featuring an internal hardware-level prefetcher that hides storage overhead by moving data from SSDs into an integrated DRAM cache. Furthermore, we provide a user-level prefetcher API that allows LLM inference frameworks to explicitly trigger and control these transfers, enabling data delivery at near-maximum PCIe bandwidth through application-aware scheduling.

\item \textbf{Empirical Potential Analysis and FPGA Prototyping:} We validate \sysname{} by evaluating its performance potential using production-grade SK hynix CMM~\cite{skhynixcmm} and Gen5 NVMe SSDs~\cite{kioxiacd8p}, while further demonstrating functional feasibility through an FPGA-based hardware prototype. Our evaluation shows that \sysname{} achieves a $1.80\times$ throughput improvement over NVMe-oF-based disaggregated storage baselines in large-scale LLM inference.
\end{itemize}

 \section{Background and Motivation}

\subsection{Evolution of Inference Memory Hierarchy}

\begin{figure}[t]
\centering
\includegraphics[width=0.48\textwidth]{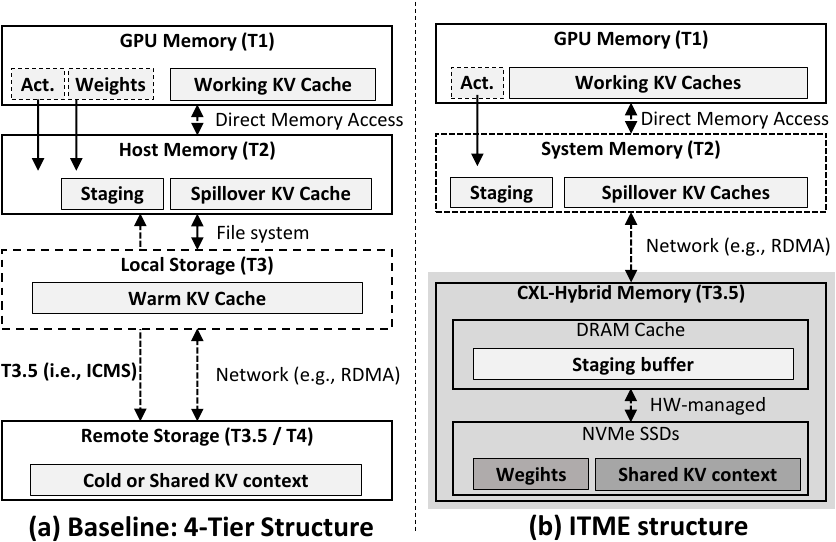}
\caption{Structural comparison of memory hierarchy}
\label{fig:itme-tier}
\end{figure}

The inference memory system is architected as a multi-tier hierarchy to balance the conflicting requirements of high-bandwidth access and high-density storage. Figure~\ref{fig:itme-tier} (a) illustrates this hierarchy, which organizes data based on its latency sensitivity and capacity footprint. GPU memory (T1) serves as the highest tier, utilizing HBM to provide the peak memory bandwidth necessary for real-time token generation. Directly supporting this is host memory (T2), which functions as a byte-addressable overflow tier for KV blocks evicted from T1. Despite their performance, the physical scaling constraints of these memory-centric tiers inevitably lead to a capacity wall as model parameters and context lengths increase. To provide deeper capacity, the hierarchy incorporates storage-based layers: single-server NVMe SSDs (T3) and cluster-wide shared storage (T4), which provide individual node capacity and massive cross-node storage, respectively.

The emergence of agentic AI and multi-turn workloads has exposed a critical capacity and accessibility gap between server-internal NVMe SSDs (T3) and remote shared storage (T4). While local storage (T3) provides fast local access, its limited physical capacity forces the system to frequently evict inference states, losing the opportunity for context reuse in subsequent turns. Conversely, while remote storage (T4) offers the necessary scale, it acts more as a cold archive rather than an active memory tier. The high latency overhead of retrieving remote states from remote storage (T4) makes it impractical to support the frequent, real-time context retrieval required by long-context inference workloads. To bridge this gap, recent industry efforts have introduced NVIDIA CMX context memory storage~\cite{nvidia_icms_2026}, a technology designed to accelerate remote access and establish a specialized shared context tier (T3.5). By providing a disaggregated, shared context layer, CMX context memory storag enables cumulative inference states, such as shared KV caches, to remain reusable and warm across the cluster. Currently, most of these implementations are DPU-centric, utilizing specialized processors like NVIDIA BlueField to offload networking and storage stacks~\cite{nvidia_icms_2026}. While DPU-based approaches improve remote-path efficiency, they require expensive, high-compute hardware and complex software stacks to manage the networking and storage layers. Instead of providing direct, low-overhead hardware access, these solutions impose a heavy management burden, demanding significant engineering effort to optimize the underlying remote storage tiers.

Our proposed \sysname{} realizes this shared context tier (T3.5) by transforming SSD-backed capacity into a direct-access memory expansion. As illustrated in Figure~\ref{fig:itme-tier} (b), \sysname{} adopts a more efficient approach by presenting itself as a remote memory server via a CXL-hybrid-memory, breaking the dependency on expensive DPUs and utilizing commodity, low-cost RNICs. The internal hardware controller directly issues NVMe requests, bypassing the traditional software-defined storage stack. To efficiently manage remote transfers, \sysname{} employs a DMA-based pipeline that orchestrates data movement, effectively masking network latency and ensuring continuous high-throughput data delivery. By transforming backing SSDs into an active, byte-addressable memory pool, \sysname{} can host TB-scale inference states and efficiently serve highly predictable long-context data to maximize CXL-hybrid-memory performance for subsequent inference turns.

\begin{figure}[t]
\centering
\includegraphics[width=0.48\textwidth]{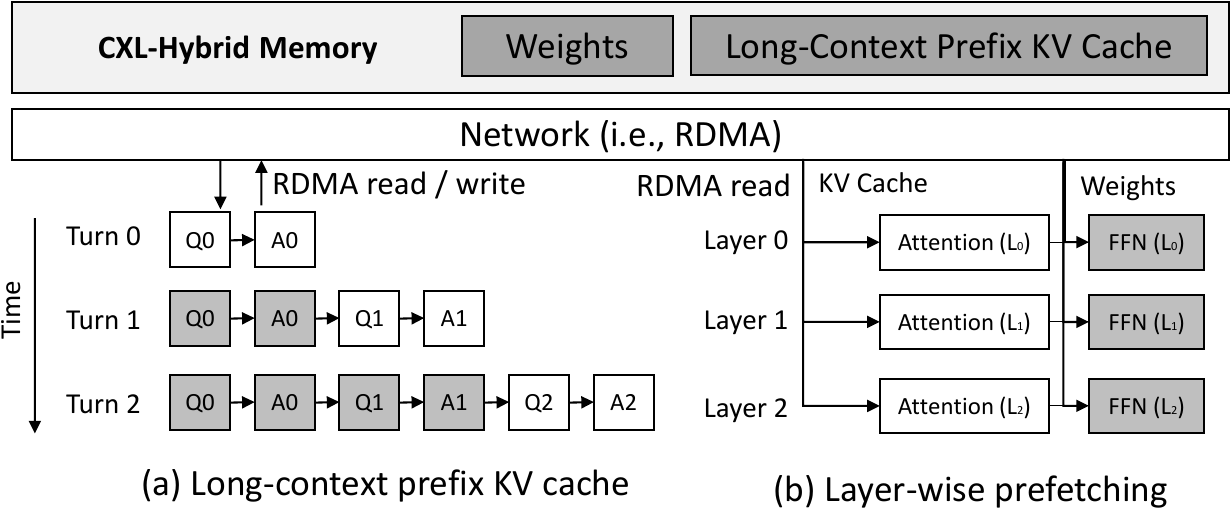}
\caption{Access behavior and prefetching opportunities for model weights and long-context prefix KV cache}
\label{fig:llm-pattern}
\end{figure}

\subsection{Exploiting Predictability in LLM Inference}
The architectural efficiency of \sysname{} stems from the high predictability of specific LLM data types (Table~\ref{tab:data_analysis}). By exploiting these deterministic access patterns, \sysname{} proactively manages data to effectively mask the latency of the remote CXL-hybrid memory.

\begin{itemize}
\item \textbf{Model Weights:} Weights are consumed in a strict layer-by-layer sequence during the forward pass. This predictability enables \sysname{} to utilize layer-wise prefetching, successfully overlapping I/O with computation. Offloading massive weight footprints (e.g., 325 GB for OPT-175B in FP16) to the CXL-hybrid memory (T3.5) reclaims valuable T1/T2 capacity for latency-sensitive active states.

\item \textbf{Long-context Prefix KV Cache:} Multi-turn and agentic workloads exhibit an append-write, sequential-restore access pattern. As shown in Figure~\ref{fig:llm-pattern} (a), \sysname{} leverages scheduler-aware fetching to restore shared-prefix KV chunks from the CXL-hybrid memorys ahead of execution. Rather than eagerly reconstructing the entire prefix on the GPU, chunks are staged and replayed layer-wise. This allows prefix reuse to overlap with attention execution while efficiently exploiting the large sequential chunk transfers of the hybrid tier.

\item \textbf{Activations and Working KV:} Due to their extreme latency sensitivity, activations and the incrementally growing working KV cache are strictly prioritized for GPU memory or host memory. Any access delay during the decoding phase causes severe pipeline stalls. Because their total memory footprint remains relatively small, GPU and host memory can comfortably satisfy their capacity demands without requiring offloading to \sysname{}.
\end{itemize}

By targeting the deterministic access patterns of weights and long-context KV caches, \sysname{} establishes an efficient prefetching pipeline from the CXL-hybrid memory to GPU memory, using host memory as an intermediate staging layer (Figure~\ref{fig:llm-pattern} (b)).

\subsection{Pipelined Multi-tier DMA-based Prefetching}
The key insight is that layer execution exposes prefetch opportunities that can overlap data movement with GPU computation. Rather than waiting until data is demanded, \sysname{} issues prefetches for upcoming blocks during the execution of earlier layers, allowing CXL-hybrid memory (T3.5)-to-GPU memory (T1) transfers to progress ahead of use. This pipelined execution model helps hide a significant fraction of transfer latency behind useful computation. Furthermore, DMA-based hardware prefetching removes the host CPU from the critical path, reducing software overhead and effectively masking the overall system latency

\begin{figure}[t]
\centering
\includegraphics[width=0.38\textwidth]{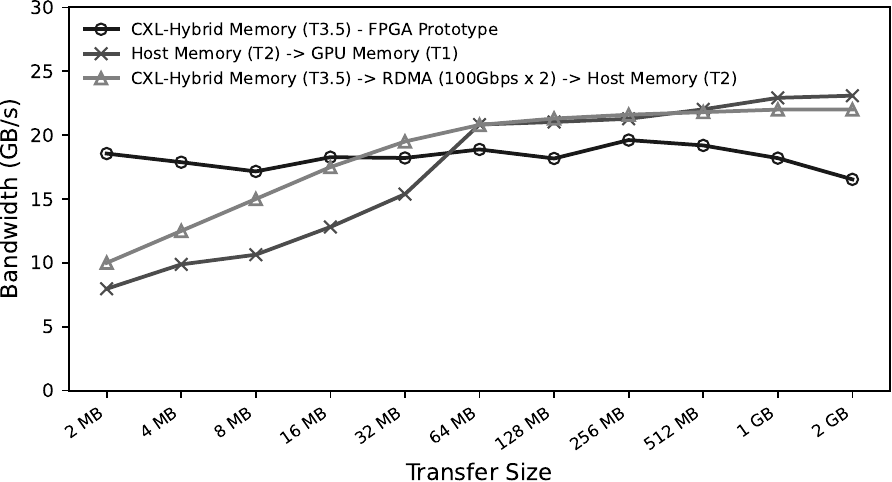}
\caption{Bandwidth (GB/s) across different DMA paths for data sizes ranging from 2 MB to 2 GB.}
\label{fig:multi-bw}
\end{figure}

As shown in Figure~\ref{fig:multi-bw}, the effective throughput across the entire data path from the CXL-hybrid memory to host memory and subsequently to GPU memory increases with transfer size and approaches the hardware limit for sufficiently large transfers. Notably, measurements from our FPGA prototype demonstrate that when data is effectively staged into the internal DRAM cache via prefetching, the CXL-hybrid memory sustains a near-peak throughput of approximately 18 GB/s across all evaluated transfer sizes. For model weights, the transfer granularity is naturally large, for example, 1344 MB for Llama-3.1 70B. This allows \sysname{} to utilize the available PCIe bandwidth efficiently. Combined with the deterministic layer-by-layer access pattern, this property enables a DMA-based pipeline in which the next layer's weights are prefetched while the current layer is still under execution. For prefix KV cache, individual blocks can be relatively small, such as 16 MB at a block size of 128. To improve transfer efficiency, \sysname{} groups multiple KV blocks into a larger aggregated transfer unit, for example, 32 blocks forming a 512MB request. Although the transfer efficiency from host to GPU memory can drop for small, non-aggregated transfers, the use of the CXL-hybrid memory together with 200Gbps RDMA enables these larger units to be streamed effectively through the pipeline. This hardware-driven data movement allows the CXL-hybrid memory to sustain high effective throughput and remain ahead of GPU demand.
\section{CXL-Hybrid Memory Architecture}

\begin{figure}[t]
\centering
\includegraphics[width=0.45\textwidth]{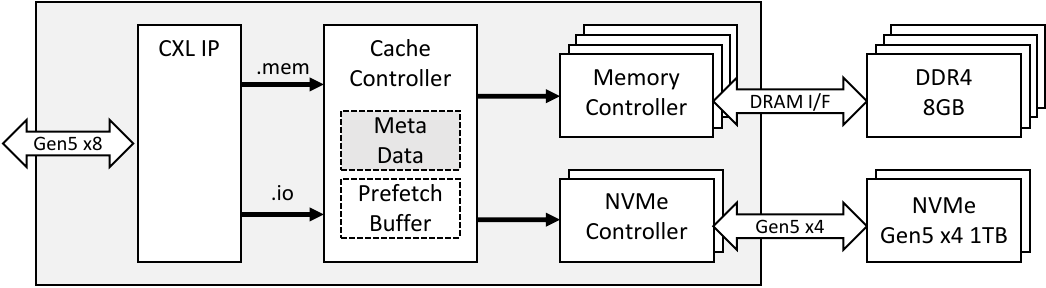}
\caption{CXL-Hybrid Memory Architecture}
\label{fig:cxl-hybrid-arch}
\end{figure}

CXL-hybrid memory is a CXL-enabled hardware device~\cite{zeng2025cxl-cmmh} designed explicitly for massive, cost-effective memory expansion. By providing cache-coherent~\cite{dirctcxlatc2022}, byte-addressable access, the CXL protocol enables the host CPU to seamlessly utilize terabyte-scale NAND flash as a massive extension of standard main memory via conventional load/store instructions.

Internally, the device achieves this massive capacity by integrating a high-density NAND flash with a hardware-managed DRAM cache. This hybrid design operates entirely transparently to the upper-level software; the DRAM cache serves as a high-speed buffer to mask backend read latencies, ensuring the expanded NAND capacity delivers efficient performance. By operating over the PCIe physical layer and exposing itself to the host operating system as a CPU-less NUMA node, the device successfully bypasses the physical pin-count limitations of traditional DDR channels~\cite{wang-cxl-realworld-arxiv24}. Consequently, CXL-hybrid memory provides legacy applications with a high-capacity, directly accessible memory space without requiring complex software-level memory management or incurring traditional storage overheads.

\subsection{Architecture and Hardware Design}
Figure~\ref{fig:cxl-hybrid-arch} illustrates the proposed CXL-hybrid memory architecture, which integrates a high-capacity NVMe SSD and a transparent, hardware-managed DRAM cache into a unified, byte-addressable memory space. A primary design objective is to fully saturate the host PCIe bandwidth to support high-speed DMA-based pipelining among GPUs, the host CPU, and RDMA NICs. Accordingly, the CXL-hybrid memory in our design can be utilized as a general-purpose memory expansion, while its performance is further maximized through specialized high-speed access optimized for the predictable data patterns targeted in this work. To achieve this, our design focuses on the following key points:

\noindent \textbf{ Hardware-Managed DRAM Cache}
To bridge the performance gap between the host interface and backend SSDs, our architecture employs the internal DRAM as a software-transparent, hardware-managed cache. An on-chip Metadata Buffer, implemented in SRAM, manages and tracks the state and location of each cache line. Each metadata entry consists of the following fields: \textbf{valid (1-bit)}, \textbf{tag (9-bit)}, \textbf{dirty (1-bit)}, and \textbf{age (4-bit)} for an LRU-based replacement policy. This metatdata is primarily managed by the following two functional units:

\begin{itemize}
\item \textbf{Hit/Miss Checker:} Upon receiving a read or write request from the host, the checker evaluates the 16-way set-associative metadata to determine a hit or miss.
\begin{itemize}
\item \textbf{Metadata Read:} It first accesses the metadata buffer to verify the \textit{valid} and \textit{tag} fields.
\item \textbf{Metadata Write (Cache Hit):} In the event of a hit, it updates the \textit{age} information to maintain the replacement policy.
\item \textbf{Dirty Update:} If the transaction is a write request, the \textit{dirty} bit is additionally updated to 1.
\item \textbf{Parallel Processing:} The system includes two hit/miss Checkers, with one dedicated to each CXL slice (data path) to ensure high-throughput request handling.
\end{itemize}

\item \textbf{Miss Line Handler:} This unit coordinates the metadata updates required during cache misses and evictions.
\begin{itemize}
    \item \textbf{Cache Miss Handling:} The handler facilitates the transfer of the requested page from the SSD to the DRAM cache and performs a \textbf{metadata write} to update the corresponding entry (e.g., setting the \textit{valid} bit from 0 to 1).
    \item \textbf{Cache Eviction:} When the cache is full, the handler identifies a victim line based on the LRU policy. It then executes a \textbf{metadata read and Write} to invalidate the evicted line (setting the \textit{valid} bit from 1 to 0) before staging the new data.
\end{itemize}
\end{itemize}

To ensure data consistency among these concurrent units, the metadata interface controller implements a hardware-level locking mechanism. Every metadata update is executed as a three-stage atomic operation: (1) metadata read and lock, (2) metadata Write, and (3) metadata Unlock. This sequential process requires three clock cycles per update, which can limit the metadata throughput during intensive processing. This internal logic and the resulting metadata footprint are the primary factors in determining the optimal DRAM and SRAM capacity for the system.

\begin{table}[t]
\centering
\caption{Metadata (SRAM) requirements}
\label{tab:metadata-req}
\resizebox{\columnwidth}{!}{
\begin{tabular}{ccccc}
\toprule
\textbf{DRAM Size} & \textbf{SSD Size} & \textbf{Index Bits} & \textbf{Tag Bits} & \textbf{Metadata (SRAM)} \\ \midrule
32 GB & 1 TB (1-ch) & 19-bit & 9-bit & 15 MB \\
32 GB & 2 TB (2-ch) & 19-bit & 10-bit & 16 MB \\
64 GB & 1 TB (1-ch) & 20-bit & 8-bit & 28 MB \\ \bottomrule
\end{tabular}
}
\end{table}

\noindent \textbf{DRAM Capacity and Metadata (SRAM) Management.}
While scaling up the storage and cache capacities can improve the hit rate, it requires a careful analysis of the on-chip SRAM overhead used for the metadata buffer. In our architecture, the metadata capacity is determined based on a 4~KB cache line and 16-way set associativity. As summarized in Table~\ref{tab:metadata-req}, the SRAM requirement exhibits a non-linear relationship with the DRAM cache size. For a baseline configuration of a 32~GB DRAM cache and a 1~TB SSD, the system manages 512~K sets, which translates to a 19-bit index and a 9-bit tag. The total metadata, comprising valid, dirty, tag, and age bit, amounts to 15~MB.

\begin{itemize}
\item \textbf{SSD Capacity Scaling:} Doubling the SSD capacity from 1~TB to 2~TB (2-ch) while keeping the DRAM at 32~GB requires only one additional tag bit (10-bit tag). This results in a marginal 6.7\% increase in the SRAM footprint, bringing it to 16~MB.
\item \textbf{DRAM Capacity Scaling:} Conversely, doubling the DRAM cache size to 64~GB for a 1~TB SSD doubles the number of cache sets to 1~M, requiring an expanded 20-bit index. This leads to a disproportionate 86.7\% surge in SRAM requirements, pushing the footprint up to 28~MB.
\end{itemize}

Considering the significant SRAM penalty, expanding the DRAM to 64~GB is not resource-efficient. Thus, we selected 32~GB as the baseline capacity, which strikes a practical balance for our hardware-managed cache.

\begin{table}[t]
\centering
\caption{Bandwidth of Internal Channel Configurations}
\label{tab:channel-exp}
\resizebox{\columnwidth}{!}{%
\begin{tabular}{cccccc}
\toprule
\textbf{DRAM} & \textbf{SSD} & \multicolumn{2}{c}{\textbf{Host $\leftrightarrow$ DRAM (GB/s)}} & \multicolumn{2}{c}{\textbf{SSD $\leftrightarrow$ DRAM (GB/s)}} \\
\cmidrule(r){3-4} \cmidrule(l){5-6}
\textbf{Ch.} & \textbf{Ch.} & Ideal & Meas. & Ideal & Meas. \\
\midrule
2-ch & 1-ch & 15.0 & 10.0 & 12.0 & 9.0 \\
4-ch & 1-ch & 23.3 & 18.0 & 12.0 & 9.0 \\
\textbf{4-ch} & \textbf{2-ch} & \textbf{23.3} & \textbf{18.0} & \textbf{24.0} & \textbf{18.0} \\
\bottomrule
\end{tabular}%
}
\end{table}

\noindent \textbf{Hardware Topology and Bandwidth Alignment.}
With the baseline configuration of 32~GB DRAM and a 2-channel SSD established to mitigate the SRAM overhead, our subsequent design focus shifts to ensuring that the internal data paths do not bottleneck the overall system performance. To achieve full line-rate utilization of high-speed host interconnects (e.g., PCIe x8), the system must balance the throughput across two critical paths: the Host-to-DRAM path for direct CPU access and the SSD-to-DRAM path for backend data staging.

As summarized in Table~\ref{tab:channel-exp}, our empirical evaluation demonstrates the effectiveness of the proposed memory configuration. A baseline setup utilizing 2-channel DRAM and a 1-channel SSD achieves only 10~GB/s for Host-to-DRAM transfers, falling short of the peak interface potential. Upgrading the memory to a 4-channel DRAM configuration (our 32~GB setup) improves the Host-to-DRAM throughput to an optimal 18~GB/s. However, if this is paired with only a 1-channel SSD, the backend SSD-to-DRAM throughput remains severely constrained at 9~GB/s. This bandwidth mismatch is critical, as it starves the hardware prefetcher and limits its ability to supply data to the cache at a rate sufficient to saturate the host-side bandwidth during cache misses.

By finalizing the topology with both 4-channel DRAM and a 2-channel SSD, we effectively resolve this bandwidth mismatch. This balanced configuration enables the system to refill the DRAM cache at the same rate the host accesses it, consequently enabling highly effective hardware prefetching. Note that the marginal gap between the theoretical ideal bandwidth and our measured throughput is primarily attributed to the 3-stage metadata updating overhead discussed earlier. Ultimately, this design not only satisfies the metadata footprint constraints but also aligns the internal memory performance closely with the peak capabilities of the host interface (e.g., PCIe x8), facilitating consistent, high-speed data delivery.

\noindent \textbf{SSD Durability and Write Mitigation.} Since CXL-hybrid memory integrates NAND flash, managing SSD durability is essential for long-term reliability. We leverage data-specific access patterns to minimize the wear-leveling impact of frequent LLM inference cycles. Read-dominant model weights occupy the bulk of the storage capacity ($\sim$105 GB for Llama-3.1 70B) but remain static once loaded into the CXL-hybrid memory. By managing these parameters in a dedicated read-only region of the SSD, the system eliminates write amplification for the largest data structures. Prefix KV cache for sequential append addresses the challenges of dynamic data by exploiting the incremental nature of token generation. Instead of fragmented updates, the system stores new KV tokens using sequential-append writes, which aligns with the internal flash management of the NVMe controller. This approach transforms KV cache management into large, block-aligned sequential streams, significantly reducing the write amplification factor (WAF) and minimizing internal garbage collection overhead to sustain high backend throughput~\cite{hu2009write}.

\subsection{User-Directed Prefetching Interface}
While the previously described topology aligns internal data paths for consistent throughput, providing a specialized prefetch API at the user level enables direct, high-performance data paths that bypass hardware-level LRU management and associated tracking overhead. To leverage this, we provide a user-level library that exposes explicit software control over the internal prefetch engine. This interface allows applications to proactively stage data into the DRAM cache, effectively masking SSD latency by exploiting the deterministic access patterns of LLM workloads.

\begin{figure}[t]
\centering
\includegraphics[width=0.45\textwidth]{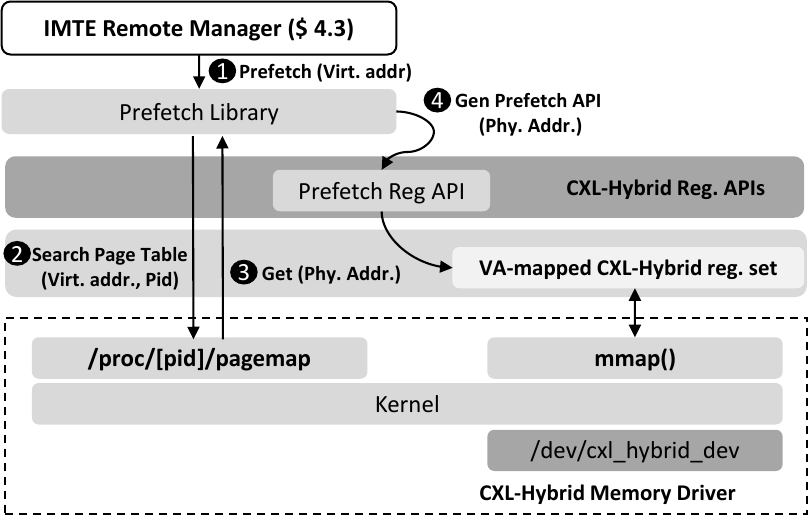}
\caption{Example Walkthrough  of User-Directed Prefetching}
\label{fig:prefetch-example}
\end{figure}

\noindent \textbf{Example Walkthrough and Implementation.} 
As illustrated in Figure~\ref{fig:prefetch-example}, the prefetch walkthrough translates an application-level virtual address (VA) \ding{202} into a hardware-compatible command. The library first performs a page table lookup through the kernel's \textit{/proc/[pid]/pagemap} interface \ding{203} to obtain the corresponding physical address (PA) \ding{204}. This PA, together with the required page count, is then written to the VA-mapped CXL-Hybrid register set \ding{205}, which is exposed to user space via \textit{mmap()}. This direct access path triggers the hardware engine to stage the specified pages from the SSD into the DRAM cache with minimal software intervention.

The library provides three primary APIs tailored to different data granularities. \textit{chm\_prefetch(addr)} targets a single 4 KB page and can optionally apply an alignment check to skip unaligned addresses and avoid redundant hardware commands. For structured data, \textit{chm\_prefetch\_object(ptr, size)} automatically determines and prefetches the page range covering a specific memory object, such as a tensor. Finally, \textit{chm\_prefetch\_size(addr, count)} reduces software overhead for large deterministic streams by fetching $N$ contiguous pages from a base address.

\noindent \textbf{Control Logic and Operational Overhead.} Prefetch commands are inserted into an internal hardware FIFO queue, where the controller uses miss status holding registers (MSHRs) to overlap multiple outstanding SSD reads. This hardware-level parallelism masks backend storage latency by staging data into the DRAM cache before it is requested. Our measurements indicate that the operational overhead of \textit{chm\_prefetch()} is approximately 3~$\mu$s, primarily due to VA-to-PA translation through \textit{pagemap}. For deterministic workloads in which weights are directly mapped via \textit{mmap()}, this overhead drops to about 1~$\mu$s. Given that a single command triggers the movement of a 1~MB chunk, this 1--3~$\mu$s latency accounts for less than 0.1\% of the total transfer time, making the software-side intervention negligible. Furthermore, because the prefetch engine operates independently, the system can overlap current-layer computation with next-layer weight staging, sustaining high throughput even when model parameters exceed the DRAM capacity of the CXL-hybrid memory.
\section{\sysname{} Implementation}

\begin{figure}[t]
\centering
\includegraphics[width=0.45\textwidth]{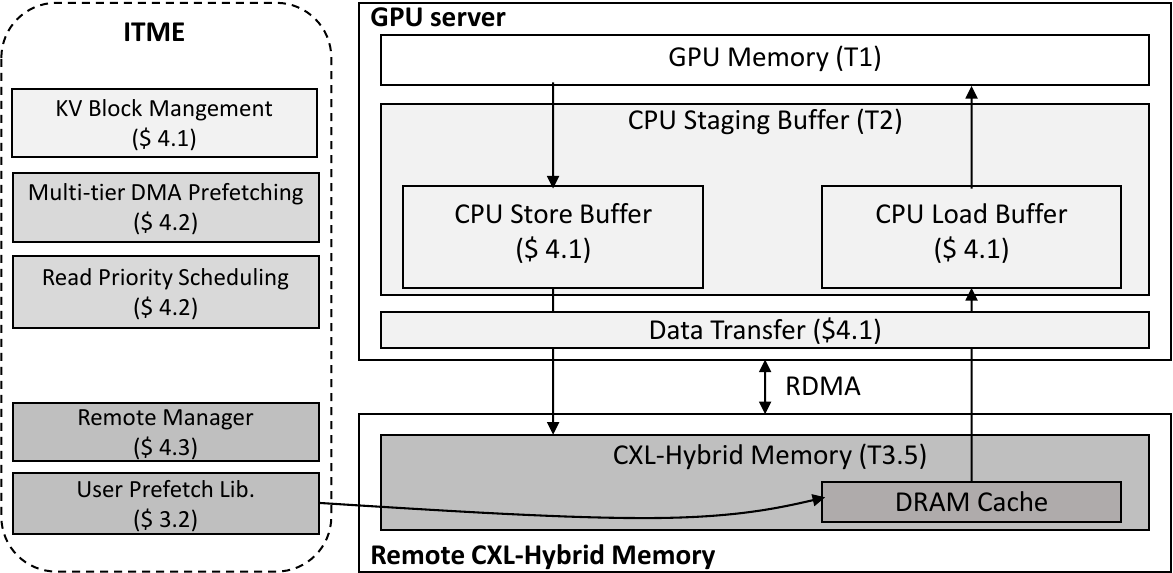}
\caption{Overall Architecture of \sysname{}}
\label{fig:itme-overall}
\end{figure}

In this section, we present the overall architecture of our proposed \sysname{}, implemented on top of the vLLM framework~\cite{kwon2023vllm}, across the multi-tier memory hierarchy, as illustrated in Figure~\ref{fig:itme-overall}.

\subsection{KV Cache Block Management}
\noindent \textbf{CPU Staging Buffers: Store and Load Buffer.}
\sysname{} allocates two dedicated pinned-memory~\cite{nvidia_cuda_prog_guide_pinned} staging buffers to isolate GPU execution from the access latencies of remote CXL-hybrid memory. This architecture effectively decouples GPU computation from the underlying storage I/O, allowing the GPU to maintain peak utilization while data movement is handled asynchronously. These buffers are specialized for bidirectional data flow: an eviction buffer for writes and a load ring buffer for reads. On the eviction path, the buffer gathers and organizes evicted blocks into large-scale chunks (e.g., 512,MB) managed in a queue format. This queue-based chunking not only optimizes high-bandwidth transfers but also naturally absorbs outgoing write traffic. To match this coarse granularity, the load ring buffer stages prefetched chunks in a fixed-depth circular queue. By maintaining a lead of 2--3 chunks, it enables the prefetch thread to consistently stay ahead of the GPU’s computational demand, facilitating a continuous and high-bandwidth data flow.

The total capacity of these staging buffers is configurable as a proportion of available host memory, offering a flexible trade-off between I/O resilience and memory footprint. While a larger staging buffer can better absorb transient burst traffic and protect the read-priority pipeline (Section 4.2), allocating extensive pinned memory naturally increases host-side resource overhead. This tunability allows \sysname{} to adapt to diverse hardware environments, balancing robust I/O performance against specific system memory constraints.

\noindent \textbf{GPU Eviction Block Management.} 
Efficient multi-tier KV caching relies on maintaining data contiguity across storage boundaries to enable high-bandwidth prefetching without reorganization overhead. To maximize offloading performance, \sysname{} aggregates evicted KV cache blocks into large-scale chunks (e.g., 512 MB) within the CPU staging buffer. During the eviction process, \sysname{} sequentially appends blocks into these chunks as they are released from GPU memory. This sequential appending naturally restores the logical sequence of KV blocks, providing a unique opportunity during the prefill phase, which accounts for the bulk of the stored data~\cite{zhong2024distserve}. Since blocks from a single request are freed simultaneously and share the same LRU, they are captured as a coherent, contiguous group within the staging chunk.

This data organization is designed to exploit the high-bandwidth sequential write capabilities of RDMA and remote CXL-hybrid memory. Consequently, when these request-related blocks are later prefetched, they can be read as a large, contiguous stream, enabling peak sequential throughput from remote CXL-hybrid memory. While decode-phase blocks are naturally interleaved across concurrent requests, \sysname{} uses a configurable block size ranging from 64 to 256 tokens. As shown in Figure~\ref{fig:multi-bw}, blocks at these scales (e.g., up to 32 MB~\cite{cheng2025lmcache}) are sufficiently large to achieve reasonable I/O performance even when processed individually.

\subsection{\sysname{} I/O Scheduling and Prefetching}
The primary role of \sysname{} is to efficiently orchestrate KV cache movement across the multi-tier hierarchy, leveraging specific access patterns and device characteristics. It manages this entire lifecycle, from migrating blocks immediately upon GPU eviction to proactively prefetching them based on upcoming execution demands.

\noindent \textbf{Read-Priority I/O Scheduling.}
The physical contiguity of the buffer enables a single RDMA Write call per chunk, bypassing redundant copies and minimizing CPU overhead. These bulk transfers are executed asynchronously to prevent pipeline stalls.

However, because the remote CXL-hybrid memory relies on NVMe SSDs, unregulated asynchronous writes can severely degrade prefetch performance. A critical bottleneck arises when background SSD writes, which flush these large sealed chunks, contend for internal resources with time-critical reads. Due to the internal architecture of NVMe SSDs, if the controller’s write buffer is saturated by bulk background flushes, it creates a massive I/O bottleneck that stalls the read queue~\cite{kimatc2019}, significantly increasing retrieval latency. This read/write contention throttles the effective read bandwidth, leading to unpredictable latency excursions as the NAND interface is saturated by ongoing write traffic. 

To mitigate this bottleneck, \sysname{} implements a read-priority I/O scheduling policy that heavily regulates background writes. The scheduler enforces absolute priority for read requests. Whenever a slot in the load ring buffer becomes available, triggering an immediate data fetch, the scheduler suppresses any pending writes. The I/O slot is yielded to the read operation, and the write is re-enqueued to the CPU staging buffer, which safely absorbs the I/O backpressure. Since evicted data can be reconstructed, writes do not require immediate persistence. This flexibility allows the system to defer or drop non-critical writes during I/O contention, providing subsequent opportunities to persist the blocks in following inference turns. Furthermore, while eviction data is managed collectively in large chunks, the scheduler issues the actual writes to the remote SSD in smaller, regulated units to prevent long-tail blocking.

Consequently, \sysname{} strategically schedules these asynchronous writes during the decode phase. Although layer-wise weight reads and next-turn KV prefetching still occur during this time, the decode phase inherently provides valuable opportunistic I/O windows with relatively lower storage utilization. By interleaving the regulated write units into these available intervals, the SSD returns to a clean state before the next compute-intensive prefill phase begins. This ensures that eviction maintenance does not interfere with critical prefetching, allowing the storage tier to consistently deliver peak read bandwidth.


\noindent \textbf{Multi-Tier DMA Prefetching.} Relying on on-demand data retrieval for GPU execution incurs severe stalls due to the remote access latency. To mask this, \sysname{} orchestrates a pipelined, multi-tier DMA prefetching mechanism that proactively migrates data from the remote CXL-hybrid memory to the GPU via host memory. This is enabled by the deterministic access patterns inherent to LLM inference. During execution, \sysname{} synchronizes data movement by optimizing the transfer granularity for each data type. Model weights are streamed layer-by-layer to match the execution flow, whereas KV cache blocks are retrieved in their original aggregated order to preserve the spatial locality established during their initial storage in large chunks. To maximize throughput, \sysname{} proactively prefetches these chunks into host memory at the onset of each inference turn, ensuring they are staged and ready for the subsequent GPU-side pipeline.

\sysname{} implements differentiated fallback policies for cache misses. While a model weight miss inevitably triggers a pipeline stall, a KV cache miss is resolved through dynamic recomputation on the GPU. This design choice avoids the inefficiency of retrieving isolated missing blocks from the remote tier, which would necessitate either suboptimal chunk-sized transfers or complex retrieval logic. Consequently, recomputing the missing KV segments directly on the GPU is more performance-efficient than waiting for high-latency remote retrieval.

\noindent \textbf{Multi-level Data Granularity}
The design of data transfer granularity in \sysname{} is a multi-tier optimization strategy intended to balance hardware-level throughput with GPU-level pipeline efficiency. To maximize the performance of the underlying CXL-hybrid memory architecture, the system manages all storage and RDMA operations at a coarse-grained chunk granularity. This approach is essential because maintaining large, sequential read and write operations saturates the internal bandwidth of the device's flash-backed memory and high-speed interconnects. Smaller granularities would trigger excessive command processing overhead and internal fragmentation, which inevitably throttle the available bandwidth of the CXL-hybrid memory.

Once chunks are staged in the CPU staging buffer, the system balances a fundamental trade-off between block-wise and layer-wise transfer granularities. While a block-wise approach enables rapid slot reclamation through a single DMA operation, it serializes execution and leaves the GPU idle until the entire block transfer completes. To align with model weight streaming and maximize pipeline efficiency, \sysname{} adopts a layer-wise prefetching strategy for the KV cache. This strategy enables a fine-grained pipeline where the DMA transfer of layer $L+1$ overlaps with the GPU computation of layer $L$, effectively minimizing GPU idle time. Although layer-wise access extends buffer occupancy, \sysname{} mitigates this by decoupling the staging cache from the CXL-hybrid memory ring buffer, sustaining high-bandwidth sequential transfers while maximizing GPU utilization through seamless computation-communication overlap.

\begin{figure}[t]
\centering
\includegraphics[width=0.45\textwidth]{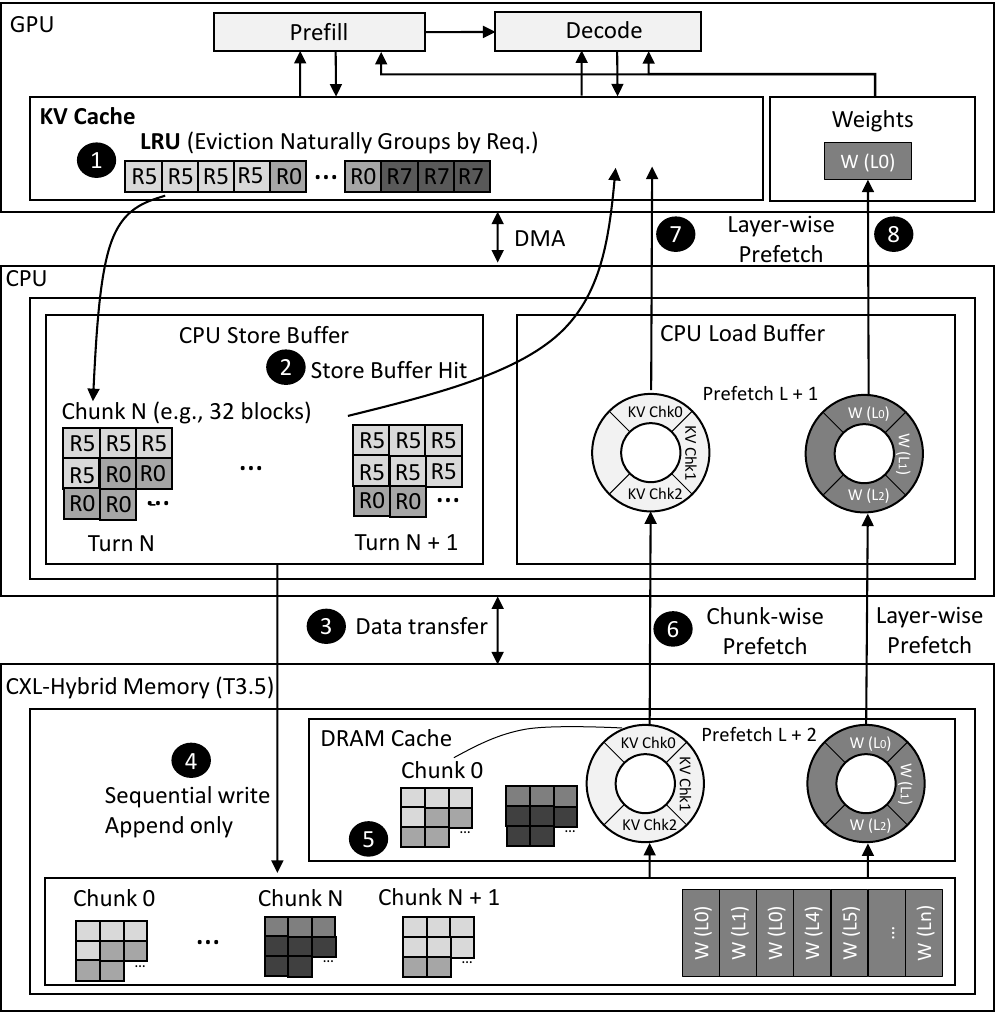}
\caption{Exmple Walkthrough of \sysname{}}
\label{fig:example}
\end{figure}

\subsection{\sysname{} Remote Manager}
The \sysname{} remote manager operates as the central orchestration layer on the CXL-hybrid memory server, specifically tasked with managing the CXL-hybrid memory device and coordinating the prefetching of KV cache chunks, model weights, and prefix caches to decouple SSD access latency from the inference critical path. When the host GPU server offloads data in coarse-grained chunks, the manager treats these transfers as structured memory objects and assigns sequential identifiers to each, establishing a direct mapping between the sequential execution of inference turns and the physical storage layout. For model weights, which remain static and are accessed repeatedly, the manager persists them in the remote tier during the initial setup phase to enable permanent reuse across multiple inference sessions without redundant transfers.

A key feature of the remote manager is its ability to trigger the device’s internal hardware prefetch engine via user-level prefetch APIs for CXL-hybrid memory (see Section 3.2). This proactive management is synchronized with the LLM inference cycle. During the decode phase of turn $N$, the manager identifies the data requirements for turn $N+1$. Since the prompt prefix for the subsequent turn is derived from the tokens generated in preceding turns, the manager retrieves the required KV cache chunks and cached prefix segments in the exact sequential order they were previously stored. In contrast, model weights are of uniform size and are managed through a circular buffer strategy. Once an RDMA read for a specific weight layer completes, the Remote Manager immediately overwrites that location by prefetching the next layers. 

The manager issues asynchronous prefetch commands to the storage controller. An internal prefetch command queue enables a continuous stream of data migration from NAND flash to the CXL-DRAM cache. This pipelined approach stages all required data in DRAM before the host issues an RDMA Read request, effectively masking SSD access latencies.

\subsection{Example Walkthrough}
Figure~\ref{fig:example} illustrates the end-to-end data movement across the GPU, CPU, and CXL-hybrid memory hierarchy. The pipeline initiates as the GPU local KV cache reaches capacity, triggering an eviction of blocks such as $R5$ and $R0$ to the CPU store buffer via asynchronous DMA (\ding{202}). When these blocks are not immediately required, they are staged in the CPU staging buffer and aggregated into coarse-grained chunks (e.g., 512,MB) to facilitate high-bandwidth sequential writes. If the GPU requests recently evicted data while it still resides in this staging area, a store buffer hit occurs (\ding{203}), allowing \sysname{} to function similarly to a conventional CPU-offload system by providing low-latency retrieval.

The read-priority I/O scheduling orchestrates the transfer of these chunks to the remote CXL-hybrid memory (\ding{204}). To prevent performance interference with latency-sensitive operations, the scheduler strategically throttles this write path during high-priority read phases. Instead, it utilizes the idle intervals of the decode phase to perform append-only updates (\ding{205}) to the CXL-hybrid memory. This prioritized approach allows the system to maximize read throughput for the inference critical path by deferring background data persistence to periods of low I/O contention.

As the next inference turn approaches, the multi-tier DMA prefetching initiates the read path. During the decode phase of the current turn, the remote manager identifies the required sequence of chunks generated in previous turns. The manager then invokes the device internal engine to perform a uhardware-level prefetch (\ding{206}), promoting the associated chunks into the internal DRAM cache. Simultaneously, the scheduler orchestrates a multi-tier transfer by initiating chunk-wise RDMA read operations (\ding{207}) to move the aggregated data from the CXL-hybrid memory to the CPU load ring buffer. Finally, \sysname{} operates a pipelined transfer that proactively pushes required data to the GPU memory ahead of each execution step. Specifically, \sysname{} employs layer-wise prefetching for both model weights and the prefix KV cache (\ding{208}, \ding{209}). By overlapping these retrievals with ongoing GPU computation, \sysname{} effectively masks the latency overhead associated with accessing the CXL-hybrid memory.

\section{Evaluation Methodology}

\begin{figure}[t]
\centering
\includegraphics[width=0.40\textwidth]{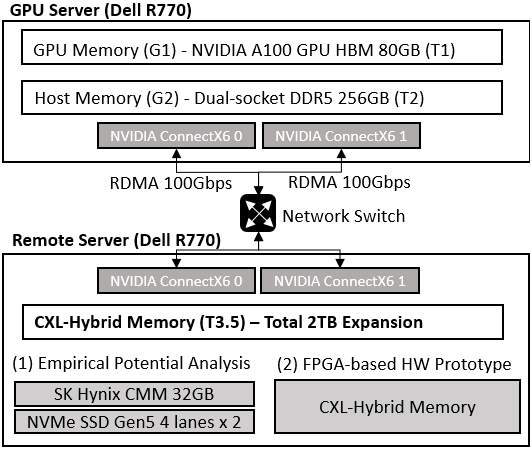}
\caption {Overview of the multi-node evaluation testbed: (a) GPU server (T1/T2) and remote CXL-hybrid memory server (T3.5) hardware configuration}
\label{fig:itme-setup}
\end{figure}

Figure~\ref{fig:itme-setup} illustrates the evaluation environment, consisting of a host and a remote CXL-hybrid memory server implemented on Dell PowerEdge R770 platforms. The host features dual Intel Xeon 6730 CPUs and 256 GB of DDR5 DRAM (128 GB per socket), representing the host memory tier (T2). It is equipped with an NVIDIA A100 (80 GB HBM) GPU (T1). The servers are interconnected via dual Mellanox ConnectX-6 100 Gbps NICs. This interconnect supports the data path requirements for the disaggregated memory hierarchy.

A comprehensive evaluation approach, utilizing two specialized environments, validates both the functional feasibility and the performance potential of the CXL-hybrid memory.

\noindent \textbf{Functional Prototyping via FPGA.} A functional prototype, implemented on an Intel Agilex 7 I-Series FPGA~\cite{altera_agilex7_i_series}, validates the hardware-level control logic and the software prefetcher. This platform integrates 32 GB of DDR4 DRAM as a hardware-managed cache for two SK Hynix Platinum P51 PCIe Gen5 NVMe SSDs~\cite{techpowerup_skhynix_p51}. This setup establishes the feasibility of our expansion tier, successfully realizing a multi-terabyte, byte-addressable memory footprint by leveraging the cost-efficiency of NAND flash.

\noindent \textbf{Performance Potential Analysis via CMM.} To evaluate the performance potential of the proposed architecture, we developed a production-grade empirical platform integrating an SK Hynix CMM~\cite{skhynixcmm} and two KIOXIA PCIe Gen5 NVMe SSDs~\cite{kioxiacd8p}. To maintain architectural consistency with the FPGA prototype, we utilized 32 GB of the CMM capacity as the DRAM cache. Leveraging the SPDK framework~\cite{intel_spdk} with 8 PCIe lanes (4 lanes per SSD), this setup achieves a peak read throughput of approximately 22 GB/s. Unlike fixed FPGA logic, this representative platform offers high degrees of freedom for prefetcher implementation. By utilizing host-side multi-threading and user-space polling, the system executes complex asynchronous I/O with high concurrency, characterizing the full performance potential of the CXL-hybrid memory (T3.5).

\subsection{Baselines and Benchmarks}
To demonstrate the advantages of \sysname{}, we implement and evaluate it within a highly optimized version of vLLM (v0.17.0)~\cite{kwon2023vllm}.

\noindent \textbf{Baseline (vLLM with CPU Offloading).} By default, vLLM handles host memory pressure by evicting and recomputing KV caches, which incurs significant overhead. To create a competitive and realistic baseline, we developed vLLM, an extended version of vLLM with the following custom implementations:

\begin{itemize}
\item \textbf{NVMe-oF Integration:} We integrated an NVMe-oF (NVMe over Fabrics) stack to enable storage access. This allows the baseline to offload and retrieve KV cache blocks from a storage node, rather than relying solely on local CPU memory or costly recomputation.
\item \textbf{Weight Prefetching:} To ensure a fair comparison with our hardware-assisted prefetching, this baseline is configured to overlap weight transfers from host memory to GPU memory with active computation. This represents an optimized software-based offloading system that minimizes the performance impact of model weights.
\end{itemize}

\noindent \textbf{Models and Datasets.} We evaluate the system using Llama-3.1 8B and 70B models to demonstrate the scalability of our architecture across different model scales. The evaluation utilizes the ShareGPT dataset~\cite{kwon2023vllm} to construct multi-turn conversation workloads with diverse sequence lengths. Furthermore, we conduct a specialized case study using the Mooncake dataset~\cite{qin2025mooncake} to analyze the system's behavior. These datasets enable us to measure the system's effectiveness in managing the cumulative growth of the KV cache footprint and its ability to mask retrieval latencies from the CXL-hybrid memory (T3.5) tier through software prefetching.

\section{Evaluation}

\begin{figure}[t]
\centering
\includegraphics[width=0.45\textwidth]{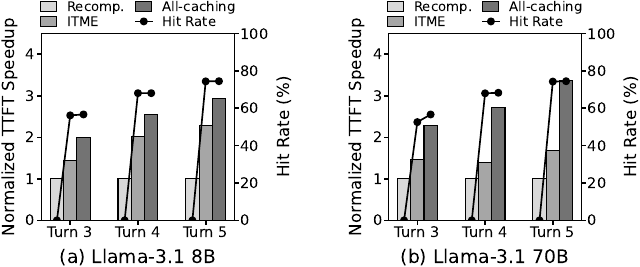}
\caption{Recomputation is set to 16 GB for 8B and 32 GB for 70B models, while All-caching consumes 80 GB of GPU memory. In contrast, \sysname{} maintains a stable 16,GB footprint by offloading 10 GB to the Host/CXL staging.}
\label{fig:overall-result-gpu}
\end{figure}

\subsection{Performance: GPU Memory (T1) vs. \sysname{}}

The performance of \sysname{} is evaluated using the ShareGPT dataset with a configuration of 128 concurrent conversations, each spanning up to 5 turns with a minimum of 2000 tokens per conversation. As memory pressure intensifies in later stages, we focus on Turns 3--5, where the total KV cache footprint reaches 40 GB.

Figure~\ref{fig:overall-result-gpu} illustrates the time to first token (TTFT) speedup for the Llama-3.1 8B and 70B models, normalized against a GPU memory (T1) baseline utilizing recomputation. To ensure a fair evaluation, both the baselines and \sysname{} are configured with weight offloading and prefetching enabled. To establish the upper performance bound of our architecture, we include an ideal GPU memory configuration with 80 GB, where the entire KV cache footprint resides within the local high-bandwidth memory.

The experimental results characterize the performance positioning of \sysname{} between high-cost local GPU memory and high-overhead recomputation. While the Ideal GPU memory configuration achieves a maximum speedup of $3.02\times$ due to its inherent bandwidth advantage, \sysname{} reaches a $1.81\times$ speedup by turn 5. Although \sysname{} is constrained by the latency of remote CXL-hybrid memory compared to local GPU memory, it offers a substantial performance gain over recomputation-based baselines. This improvement is realized whenever the system can fetch KV blocks from remote CXL-hybrid memory faster than the time required for recomputation. By successfully identifying access patterns and staging KV blocks in advance, the system minimizes the overhead of remote CXL-hybrid memory and maximizes inference efficiency.

\begin{figure}[t]
\centering
\includegraphics[width=0.45\textwidth]{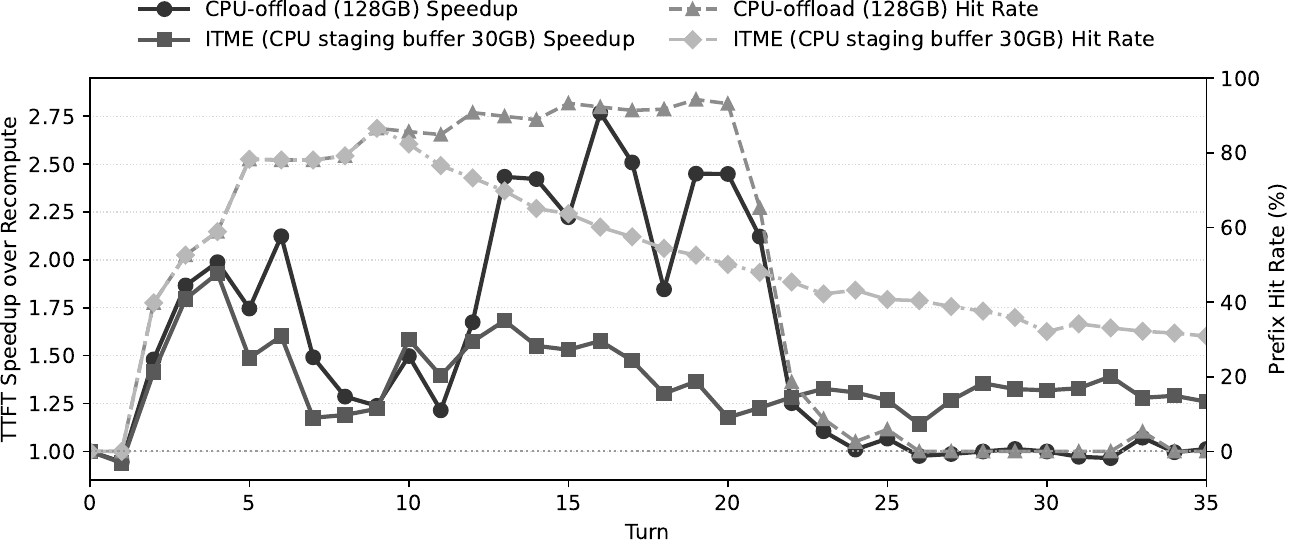}
\caption{Performance comparison between \sysname{} and CPU-offload}
\label{fig:multi-turns-comp-t2}
\end{figure}

\subsection{Performance: Host Memory (T2) vs. \sysname{}}

Figure~\ref{fig:multi-turns-comp-t2} compares the multi-turn inference performance of \sysname{} against the CPU-offload and recompute-only baselines. The figure evaluates inference efficiency using 256 concurrent conversations from the ShareGPT dataset. To stress-test KV cache capacity, we designed a 35-turn benchmark that generates a massive total footprint. In this setup, the CPU-offload baseline allocates a large 128 GB host memory (T2) for caching. In contrast, \sysname{} leverages the CXL-hybrid memory (T3.5) tier while utilizing only a fraction of host memory (T2) as a staging buffer. All results are normalized against a recomputation baseline to isolate the speedup enabled by \sysname{}'s efficient multi-tier memory management.

During the initial turns (turns 1--9), \sysname{} exhibits comparable performance to the CPU-offload baseline. This is because the active working set still fits within the host memory tier; thus, evicted KV blocks are served directly from the CPU staging buffer without incurring the additional latency of the remote CXL-hybrid memory. Regardless of the staging buffer size (30 GB vs. 128 GB), \sysname{}'s performance closely aligns with the CPU-offload baseline. This indicates that their underlying memory access characteristics are fundamentally identical as long as the working set is served within the host tier, before remote eviction begins. However, as the conversation extends (turns 10--20), a performance gap emerges. While the CPU-offload baseline maintains a high prefix hit rate near 90\%, \sysname{} experiences a gradual decline. As the KV cache volume expands with each turn, the increasing read/write traffic triggers I/O contention within the CXL-hybrid memory. This prevents some data chunks from reaching the CPU staging buffer in time, forcing recomputation for the delayed blocks, in contrast to the stable CPU-offload baseline.

Beyond turn 21, the CPU-offload baseline completely exhausts its 128 GB memory. With no space left to store new KV blocks, its caching system collapses, forcing the hit rate to 0\% and making it as slow as the recompute-only baseline. In contrast, \sysname{} continues to function by leveraging its massive CXL-hybrid memory. While the \sysname{} exhibits some performance fluctuations due to unpredictable I/O stalls during intense contention, our read-priority scheduling effectively mitigates these bottlenecks. Even under such pressure, retrieving data from the remote tier remains far more efficient than full recomputation. Consequently, \sysname{} achieves up to a 35.7\% throughput improvement over the CPU-offload baseline in these extended turns.

\begin{figure}[t]
\centering
\includegraphics[width=0.45\textwidth]{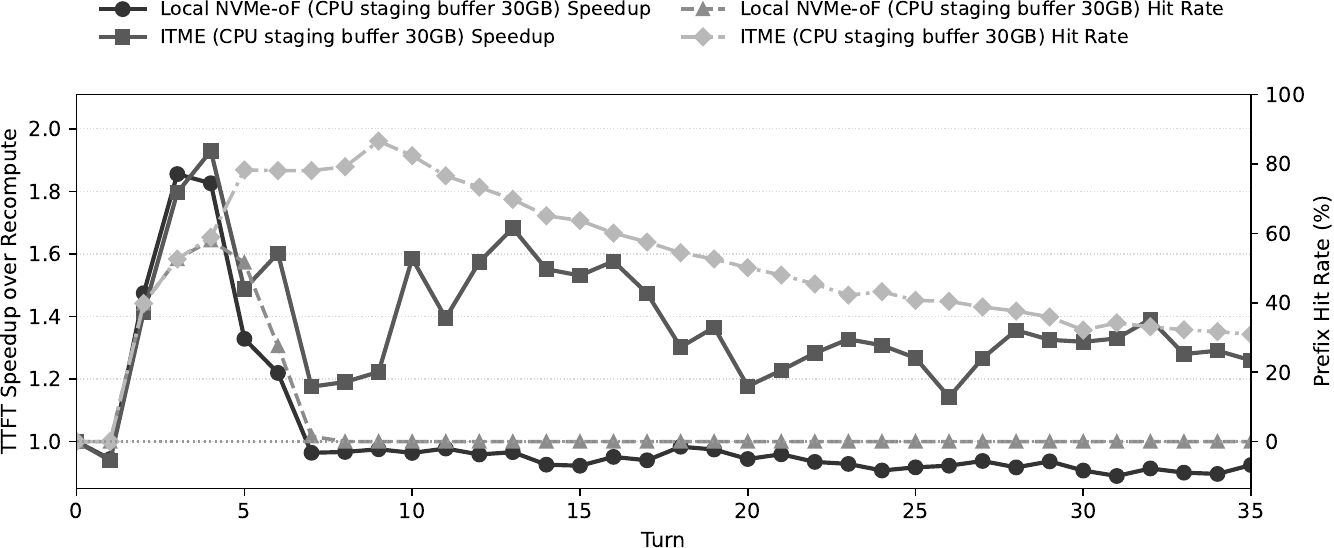}
\caption{Performance comparison between \sysname{} and an Ideal NVMe-oF baseline}
\label{fig:overall-result-nvme-of}
\end{figure}

\subsection{Performance: Local NVMe-oF (T3) vs. \sysname{}}
Figure~\ref{fig:overall-result-nvme-of} compares \sysname{} against a local NVMe-oF baseline. While real-world NVMe-oF typically operates on DPU-based JBOF systems where optimization is challenging, we utilize a local configuration to eliminate network overhead. This setup represents an ideal NVMe-oF scenario, providing a high-performance upper-bound for comparison. Both systems utilize the same CPU staging buffer and perform all data transfers in fixed-size chunks to their specific backend (either local NVMe-oF or \sysname{}). Consistent with our host memory scaling benchmarks, we utilize the same workload configuration for this comparison. To ensure a fair evaluation, both systems follow an identical caching policy: they serve KV blocks from the CPU staging buffer upon a hit, falling back to recomputation if the required data is not present in the staging buffer.

Initially, both systems exhibit identical performance as long as the KV cache remains within the 30 GB CPU staging buffer. Since requests are served directly via buffer hits, the underlying storage latency is effectively masked. However, starting from Turn 5, the performance of the Local NVMe-oF baseline degrades significantly, eventually converging with the recomputation baseline. The performance drop occurs because heavy write traffic blocks the necessary read requests. Without a way to effectively manage or delay these writes, the system cannot retrieve data while busy with incoming traffic. Furthermore, the absence of a dedicated prefetching mechanism prevents the system from hiding this delay by moving data in advance. As a result, the required KV blocks fail to reach the CPU in time, forcing the system to perform slow recomputation. While NVMe-oF could be further enhanced through specialized optimizations, our evaluation focuses on the intrinsic I/O efficiency of each backend using the same chunk-level data management. Within this scope, \sysname{} demonstrates enhanced efficiency by effectively addressing I/O bottlenecks and latency through its integrated prefetching and CXL-hybrid memory architecture.

\begin{figure}[!t]
\centering
\includegraphics[width=0.45\textwidth]{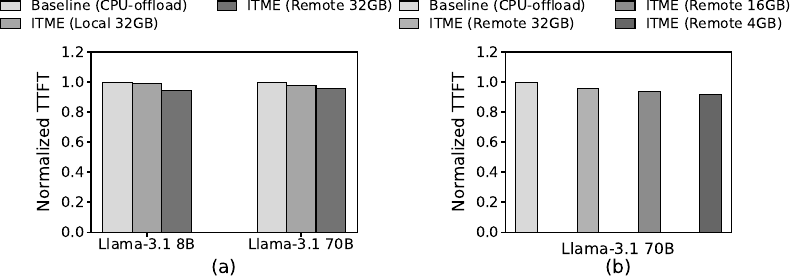}
\caption{(a) Weight Prefetching analysis; (b) Performance sensitivity to CXL-hybrid memory DRAM cache capacity}
\label{fig:weigth-prefetching-result}
\end{figure}

\subsection{Impact of Multi-tier Weight Prefetching}
To isolate the impact of weight prefetching from prefix KV cache prefetching, we evaluate the performance of \sysname{} by prefetching model weights only. We employ the Mooncake workload with a batch size of 16 and a 4K input sequence length. The primary objective of this evaluation is to demonstrate how closely \sysname{} matches the host memory baseline, where all weights reside in host DRAM. Since the baseline represents the theoretical performance upper bound, this analysis focuses on the efficiency of \sysname{} in hiding network and storage overheads through its pipelined prefetching.

Figure \ref{fig:weigth-prefetching-result} (a) compares end-to-end performance results, where the gaps between baseline, local, and remote configurations quantify storage and network overheads, respectively. While the 10.5 GB footprint of the 8B model fits within the CXL-hybrid memory DRAM cache, the 105 GB footprint of the 70B model exceeds it, yet both maintain near-baseline performance with only 1--5\% deltas. These results demonstrate that \sysname{} effectively hides storage and network latencies. Figure \ref{fig:weigth-prefetching-result} (b) evaluates the 70B model's throughput while scaling the CXL-hybrid memory DRAM capacity. Our results show that a 4GB DRAM prefetch buffer, supporting a three-depth pipeline, incurs only an 8\% performance degradation compared to the host memory baseline. While a 4 GB allocation offers a practical balance between speed and memory efficiency, it is not a complete solution. In practice, frequent model weight fetching also consumes significant read bandwidth, creating resource contention with KV cache retrieval. Therefore, achieving optimal performance requires not only proper buffer sizing but also effective scheduling to mitigate performance degradation during these simultaneous data transfers.

\begin{figure}[!t]
\centering
\includegraphics[width=0.45\textwidth]{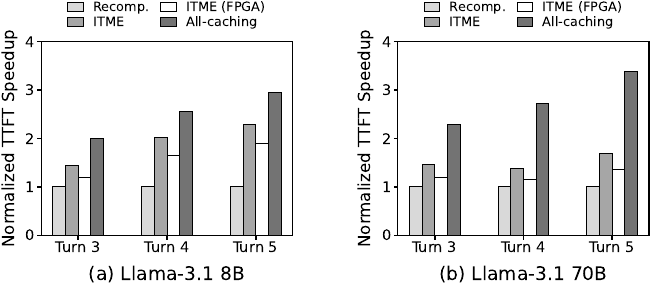}
\caption{Performance comparison including ITME (FPGA) under identical settings}
\label{fig:fpga_perf}
\end{figure}

\subsection{FPGA-based Prototype Evaluation}
Figure~\ref{fig:fpga_perf} demonstrates the hardware feasibility of the \sysname{} architecture through an FPGA-based prototype. While the CMM-based platform provides an evaluation of peak performance potential, this implementation serves to validate the hardware-level control logic and the effectiveness of the software prefetcher in a real-world system environment.

\begin{figure}[!t]
\centering
\includegraphics[width=0.45\textwidth]{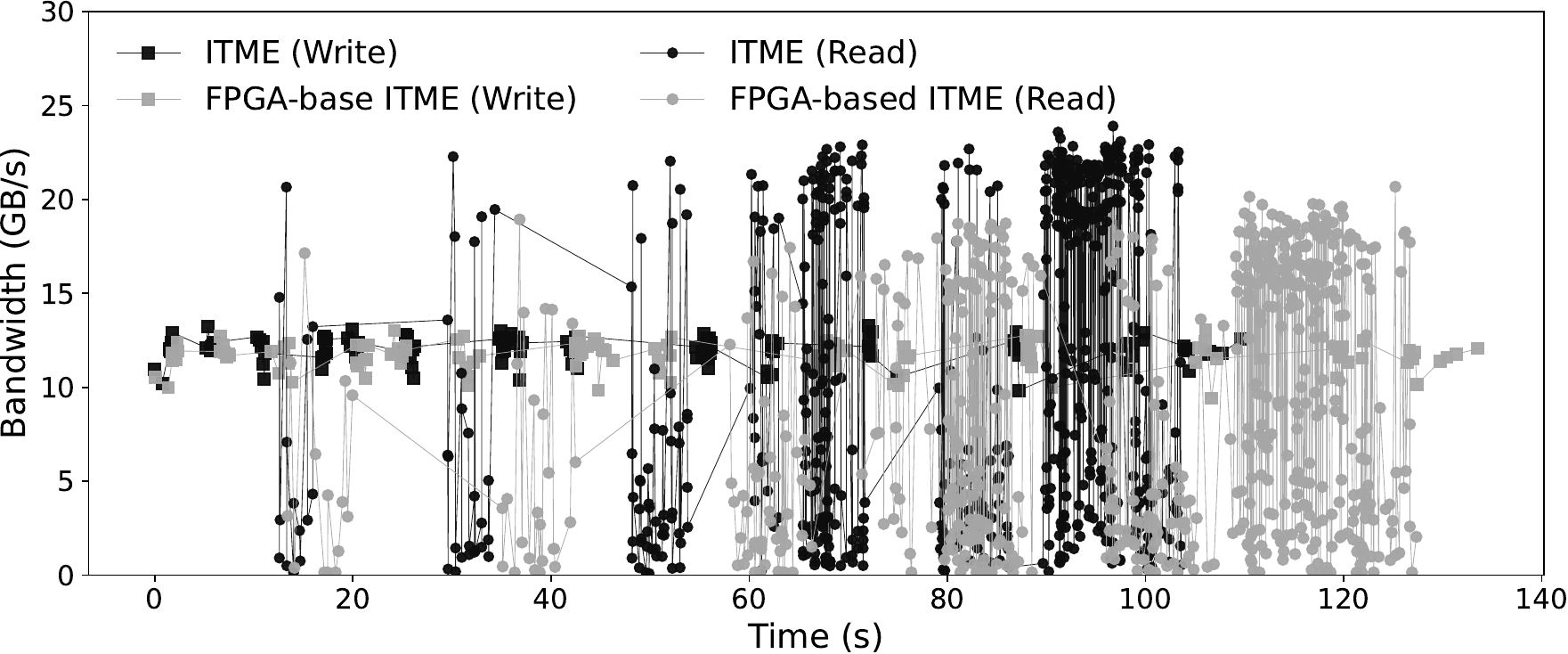}
\caption{Bandwidth comparison of the CMM based setup and FPGA prototype}
\label{fig:fpga-bw}
\end{figure}

Figure~\ref{fig:fpga-bw} compares the bandwidth of the \sysname{} CMM configuration with its FPGA prototype implementation. The FPGA prototype achieves an average of 18 GB/s for reads and 12 GB/s for writes during DRAM cache hits, representing a 20--25\% performance gap compared to the CMM-based evaluation due to hardware-level overheads. High-variance weight bandwidth is omitted for clarity, as it maintains a highly periodic and stable pattern. These results demonstrate the efficacy of our read-priority scheduling; since read operations are critical to inference latency, writes are dispatched exclusively during idle periods to minimize interference. Even if a write operation is dropped due to contention, the system ensures architectural reliability by recomputing and restoring the context in the subsequent turn. While the CMM setup remains resilient—sustaining 4--5 GB/s even under extreme contention—the current FPGA implementation's bandwidth degrades below 1 GB/s upon cache misses. We are actively optimizing the hardware logic to close this gap and reach the theoretical maximum of 23.3 GB/s.
\section{Related work}

\noindent \textbf{CXL-based Memory Expansion.} 
Compute Express Link (CXL) has emerged as a key enabling technology for memory expansion and pooling in modern datacenters. Extensive research has examined the potential of CXL-based memory for disaggregated architectures~\cite{gouk2023memory, Jang2023atc, aguilera2023memory, zhou2024lightwsp, giannoula2023daemon} and its role in tiered memory systems~\cite{Maruf2023asplos}. However, due to the limited availability of commercial CXL hardware, many studies have relied on simulations or emulations~\cite{esmaili_dokht2024mess, fridman2023cxl, arif2022exploiting}. To bridge the gap between simulation and reality, Samsung developed CMM-H~\cite{zeng2025cxl-cmmh}, an FPGA-based hybrid prototype integrating a DRAM cache and NAND flash. While it offers cost-efficient, byte-addressable capacity and persistence , its PCIe Gen4-based architecture limits available bandwidth, creating a potential bottleneck for LLM inference. In contrast, the CXL-hybrid memory in \sysname{} fully adopts the PCIe Gen5 interface, delivering the high bandwidth essential for rapid KV cache swapping in multi-tier pipelines.

\noindent \textbf{Multi-tier KV Cache and Prefetching.}
The substantial memory footprint of KV caches in LLM inference has spurred significant research into tiered memory hierarchies. Systems like FlexGen \cite{sheng2023flexgen} and DeepSpeed Inference \cite{aminabadi2022deepspeed} maximize throughput by partitioning data across GPU, CPU, and NVMe SSDs, while LLM in a Flash \cite{alizadeh2023llmflash} and PowerInfer \cite{song2023powerinfer} optimize flash-based offloading. Specialized systems such as LMCache~\cite{cheng2025lmcache} and Mooncake~\cite{qin2025mooncake} address recomputation overheads through global sharing and disaggregated architectures. Furthermore, PagedAttention \cite{kwon2023vllm}, Pensieve \cite{yu2023pensieve}, and CachedAttention \cite{gao2024cachedattention} introduce advanced multi-tier cache management and memory paging for efficient context handling. Our proposed architecture, \sysname{}, is designed to be fully compatible and synergistic with these existing software-level policies. Rather than replacing these stacks, \sysname{} serves as a high-performance memory backend that integrates seamlessly with established tiering mechanisms. Furthermore, its internal hardware-level prefetching complements various scheduling policies, effectively hiding access latency by overlapping data movement with ongoing computation. This synergy allows \sysname{} to maximize KV cache reuse and system efficiency within any established inference hierarchy.

\noindent \textbf{DPU-based Storage and KV Caching.}
The advancement of DPUs has enabled high-density storage disaggregation, leading to the development of specialized JBOF systems \cite{supermicro_jbof}. To optimize data access in these environments, several DPU-centric key-value (KV) stores have been proposed. LEED \cite{guo2023leed} and Gimbal \cite{gimbal_sigcomm21} focus on offloading storage management and indexing to the DPU's ARM cores to reduce host CPU overhead. Furthermore, software optimizations like NVMe-oF Target Offload \cite{nvmeof_target_offload_perf, nvme_offload_whitepaper} and RDMA-based caching systems such as Ditto \cite{ditto_sosp23} and FORD \cite{ford_fast22} have been introduced to bypass processing bottlenecks and achieve near-line-rate I/O.

While these prior works primarily concentrate on optimizing the I/O path between DPUs and NAND flash or improving in-memory hit rates, they often face scalability limits due to the fixed DRAM capacity and the computational overhead of DPU cores. In contrast, our system takes a different perspective by providing disaggregated CXL-hybrid memory expansion. Unlike traditional JBOF-based KV stores that rely on standard storage protocols, our approach leverages CXL-hybrid memory to offer a seamless, high-bandwidth memory tier. This enables more efficient resource utilization and addresses the memory-capacity wall in large-scale LLM inference.
\section{Conclusion}
In this paper, we propose \sysname{}, which leverages CXL-hybrid memory to provide massive, byte-addressable remote memory expansion. By exploiting the deterministic access patterns of LLM workloads, \sysname{} implements a multi-tier DMA prefetching pipeline that effectively masks storage access latencies. We validated \sysname{} using production-grade SK Hynix CMM and PCIe Gen5 SSDs, while demonstrating hardware feasibility via an FPGA-based prototype. Our evaluation confirms that \sysname{} effectively mitigates memory capacity and I/O bottlenecks through software prefetching and read-priority scheduling. Overall, \sysname{} enhances conventional CPU-offloading by providing the necessary memory expansion to accommodate large KV cache footprints beyond host memory limits, achieving up to a 35.7\% throughput improvement.

\bibliographystyle{ACM-Reference-Format}
\bibliography{ref}

\end{document}